\documentclass{article}
\usepackage{spconf,amsmath, graphicx,amssymb,bm, mathrsfs, array, multirow, algorithm, algorithmic, graphicx, cite, xcolor, stmaryrd}

\title{Convolutional Framework for Accelerated Magnetic Resonance Imaging}
\name{Shen Zhao*, Lee C. Potter*, Kiryung Lee*, Rizwan Ahmad** \thanks{Corresponding author: Rizwan Ahmad (ahmad.46@osu.edu). This work was funded by NIH R01HL135489 and NSF CCF 17-18771.}}
\address{*Department of Electrical and Computer Engineering, **Department of Biomedical Engineering, \\ The Ohio State University}
\begin{document}
%
\maketitle
\begin{abstract}
Magnetic Resonance Imaging (MRI) is a noninvasive imaging technique that provides exquisite soft-tissue contrast without using ionizing radiation. The clinical application of MRI may be limited by long data acquisition times; therefore, MR image reconstruction from highly undersampled k-space data has been an active area of research. Many works exploit rank deficiency in a Hankel data matrix to recover unobserved k-space samples; the resulting problem is non-convex, so the choice of numerical algorithm can significantly affect performance, computation, and memory. We present a simple, scalable approach called Convolutional Framework (CF). We demonstrate the feasibility and versatility of CF using measured data from 2D, 3D, and dynamic applications.

\end{abstract}
\section{Introduction}
\label{sec:intro}
MRI reconstruction from undersampled data implicitly relies on priors to provide regularizing assumptions. 
Priors in MRI reconstruction may be considered in two classes: coil properties and image structure. 
The spatial smoothness of coil sensitivity maps may be modeled as filters with finite k-space support, leading to a shift-invariant  prediction property for multi-coil k-space data. GRAPPA \cite{griswold2002generalized} uses a fully-sampled auto-calibration signal (ACS) region to solve for a linear prediction kernel and applies the kernel to recover missing k-space samples. SPIRiT \cite{lustig2010spirit} is a generalization of GRAPPA that enforces the linear predictability across the entire k-space. PRUNO\cite{zhang2011parallel} generalizes SPIRiT further by using multiple kernels satisfying the linear prediction property. The shift invariant linear prediction property may be expressed as a nullspace of a convolution operator, which is a multi-level Hankel-structured matrix. And, the finite k-space support of coil sensitivities translates to low-rank for the structured Hankel data matrix. Calibration-free methods, such as SAKE \cite{shin2014calibrationless}, employ low-rank structured matrix completion to recover missing k-space samples. To improve computational efficiency, alternatives to the Cadzow's algorithm (used in SAKE) have been proposed for single-coil \cite{ongie2017fast} and multi-coil \cite{haldar2016p} MRI. 

Sparsity in the image domain also leads to existence of approximate annihilating filters in the k-space and yields the shift-invariant linear prediction property~\cite{haldar2019linear}. Any linear filtering that sparsifies the image also yields annihilating filters in the resulting weighted k-space. In this vein, ALOHA uses a low-rank matrix completion method to recover missing samples in the weighted k-space~\cite{jin2016general}.


We present a viewpoint and computational approach, called Convolutional Framework (CF), that unifies many reconstruction techniques and provides an algorithmic approach for structured matrix completion.  Memory efficiency permits high-dimensional imaging cases not demonstrated previously. Numerical evaluations are presented for three imaging applications: 2D, 2D cine, and 3D. Discussion and summary conclude the manuscript.

\section{Methods}
We first define notation for organizing k-space data into structured arrays.
A tensor denotes a multidimensional array. Let $*$, $\circledast$, and $\boxast$ denote linear convolution, circular convolution, and valid convolution, respectively. Valid convolution maps to output points that do not depend on any boundary conditions (i.e., no padding of the input). For example, for $n$-point vector $\bm{a}$ and $m$-point vector $\bm{b}$, $m \geq n$, the lengths of $\bm{a}*\bm{b}$ and $\bm{a} \boxast \bm{b}$ are $m+n-1$ and $m-n+1$, respectively. Let $\bm{s}(\mathbb{A})$ be the row vector that lists the sizes of tensor $\mathbb{A}$. Let $\mathscr{H}_{\bm{s}(\mathbb{A})}\{ \mathbb{B} \}$ denote multi-level block Hankelization of the tensor $\mathbb{B}$ such that right multiplication of the matrix $\mathscr{H}_{\bm{s}(\mathbb{A})}\{ \mathbb{B} \}$ with vectorized $\mathbb{A}$ $(\bm{vec} \{\mathbb{A}\})$ is the vectorization of the valid convolution result between $\mathbb{B}$ and $\mathbb{A}$, i.e.,
\begin{eqnarray}
	\mathscr{H}_{\bm{s}(\mathbb{A})} \{\mathbb{B}\} \bm{vec} \{\mathbb{A}\} = \bm{vec} \{\mathbb{B} \boxast \mathbb{A}   \}.
\end{eqnarray}
Finally, $\circ$ denotes Hadamard multiplication between tensors of the same size, and $(\cdot)^H$ denotes conjugate transpose.

For generality, we denote the k-space by $\mathbb{D}$ with five dimensions: frequency encoding, first phase encoding, second phase encoding, coil, and time; dimensions are ordered and indexed by $k_x, k_y, k_z, l, t$. We choose the kernel size $\bm{s} = [f_x, f_y, f_z, N_c, f_t]$, where $N_C$ is the number of coils.

Based on the shift-invariant linear predictability assumption,  CF leverages linear prediction in all dimensions, i.e., annihilating filters exist for jointly processing all dimensions. An iterative algorithm to apply the property has two simple steps: (i) Estimate all annihilating filters from the current estimate of k-space; and (ii) Enforce annihilation for the whole k-space and update the unobserved k-space. The CF processing is summarized in Algorithm 1.
\renewcommand{\algorithmicrequire}{ \textbf{Input:}} 
\renewcommand{\algorithmicensure}{ \textbf{Output:}} 
\begin{algorithm}
\caption{Pipeline of CF }
\label{alg: 1}
\begin{algorithmic}[1]
\REQUIRE ~~\\
	$\mathbb{D}_o$: Observed k-space with zero filling\\
	$\mathbb{M}$: Sampling mask\\
	$\bm{s}$: Kernel size\\
	$r$: Rank\\
	$\text{tol}$: Tolerance
\ENSURE ~~\\
	$\hat{\mathbb{D}}^{(n)}$: Recovered k-space\\
	Initialization: $n = 0, \delta = \infty, \hat{\mathbb{D}}^{(n)} = \mathbb{D}_o$
	\WHILE{ $\delta > \text{tol}$}
	\STATE $n = n+1$
	\STATE $[\bm{\Lambda}^2, \bm{V}] = \text{EIG}(\mathscr{H}_{\bm{s}}^H (\hat{\mathbb{D}}^{(n)}) \mathscr{H}_{\bm{s}}(\hat{\mathbb{D}}^{(n)}))$
	\STATE $\bm{V} = [\bm{V}_\parallel~|~\bm{V}_\perp]$ based on $r$
	\STATE Split, reshape and flip $\bm{V}_\perp$ into kernels $\mathbb{F}_1, \mathbb{F}_2, \cdots$
	\STATE $\hat{\mathbb{D}}_u^{(n)} = \text{argmin}_{\mathbb{X}} \sum_i \| (\mathbb{D}_o + \mathbb{X})\boxast \mathbb{F}_i \|_F^2$ s.t. $\mathbb{X} \circ \mathbb{M} = 0$
	\STATE $\hat{\mathbb{D}}^{(n)} = \mathbb{D}_o + \hat{\mathbb{D}}_u^{(n)}$
	\STATE $\delta = \|\hat{\mathbb{D}}^{(n)} - \hat{\mathbb{D}}^{(n-1)} \|_F / \|\hat{\mathbb{D}}^{(n-1)} \|_F$
	\ENDWHILE
\end{algorithmic}
\end{algorithm}\\
The eigendecomposition in Step 3 extracts a null space ($\bm{V}_\perp$) of the product of two Hankel operators; this avoids a singular value decomposition of the explicit -- and very large -- convolutional matrix $\mathscr{H}_{\bm{s}}^H$. The operator product in Step 3 may be calculated with limited memory and computation using convolution with small kernels, in lieu of explicit matrices. Step 6 is a large scale least squares problem; to limit memory requirements, we avoid direct computation and instead rely on the implicit convolution operator in a gradient descent (GD) method with exact linear search (ELS). The memory requirement for GD + ELS is approximately only the original data size, in contrast to explicit construction of Hankel matrices.  Step 6 minimizes null space energy while simultaneously preserving Hankel structure and data
consistency. In contrast, SAKE enforces the Hankel structure, rank deficiency, and data consistency as three separate projections. Also, spatial, coil, and time dimensions are jointly incorporated in CF. 

If there is an ACS region, we may directly estimate the null space $\bm{V}_\perp$ from $\mathbb{D}_{\text{ACS}}$, then enforce it to hold for the whole k-space. This variant, which for 2D static MRI shares the assumptions with PRUNO, avoids the iterative step to estimate $\bm{V}_\perp$. Also, the type of convolution (linear, circular, or valid) may differ across k-space dimensions; for example, we may adopt circular convolution for the time dimension and valid convolution for the others.

\section{Experiments and Results}
We implement CF and compare to several existing techniques for 2D, 3D, and 2D cine (``2D+t'') imaging. For a fair comparison, we set the kernel size to be $5 k_x \times 5 k_y \times N_C$ for CF, SAKE, and ALOHA. Since P-LORAKS \cite{haldar2016p} uses a disk-shaped kernel, we choose a radius $3$ for P-LORAKS (``C'' version) which leads to a similar but slightly larger kernel including $29> 5\times 5 = 25$ k-space points per coil. The reconstruction SNR generally increases with the kernel size but so does the computation burden. The stopping tolerance, relative change in the whole k-space, is $10^{-3}$ for all methods.  A maximum of $200$ iterations is used, except for SAKE, in which case we also continue beyond $200$ iterations until the SNR of SAKE matches that of CF. This reconstruction is referred to as SAKE*. We fine-tuned rank selection for all methods with respect to the first dataset, then applied that choice of rank to all other datasets.  For other parameters, e.g., regularization parameter $\lambda$ for P-LORAKS, we employ published default values. For the recruitment and consent of human subjects used in this study, the ethical approval was given by an Internal Review Board (2005H0124) at The Ohio State University.

\begin{figure}[h!]
	\includegraphics[width = 8.4cm ]{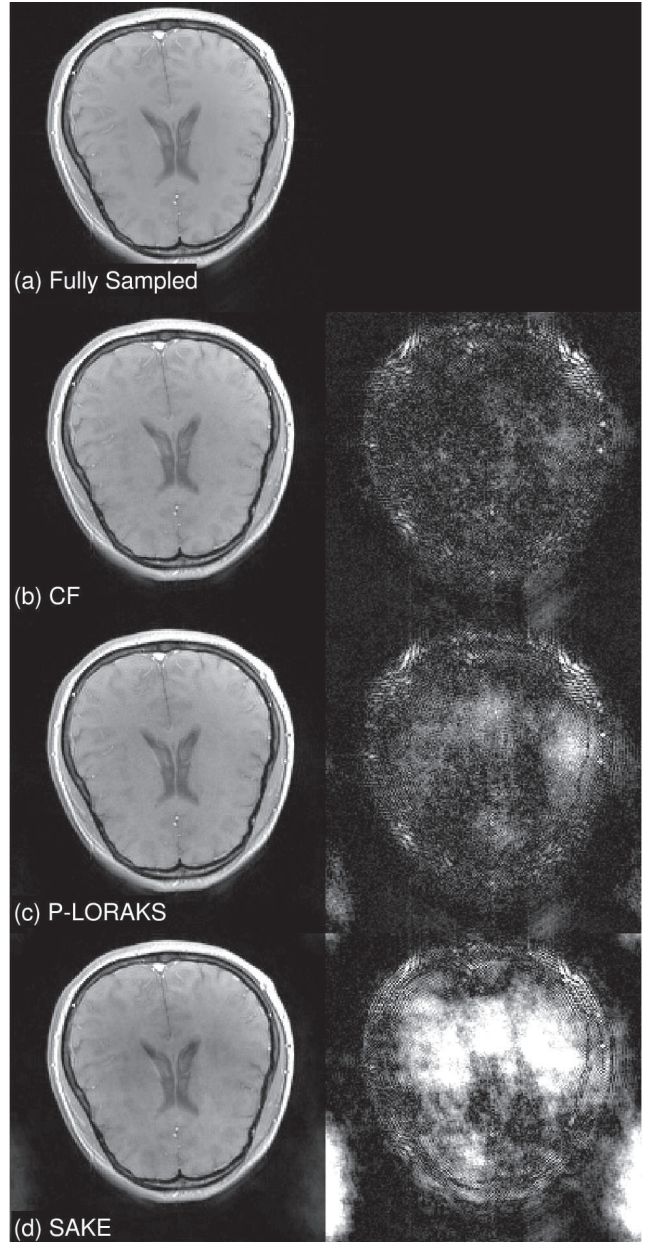}
	\centering
	\caption{A representative 2D reconstruction from dataset 2 with sampling pattern without the ACS region and acceleration rate $4$. Left: SSoS images. Right: $10\times$ absolute error.  (a) fully sampled; (b) CF ($23.0$\,dB SNR); (c)  P-LORAKS ($20.8$\,dB); (d) SAKE ($14.9$\,dB).}
	\label{Fig:1}
\end{figure}

For 2D, we retrospectively downsampled three 3T brain datasets using three different acceleration rates, R~$=4,6,$~and $8$ and four different sampling patterns: (i) 2D random; (ii) 2D random + $7k_x \times 7 k_y$ ACS region; (iii) 2D random + $17 k_x \times 17 k_y$ ACS region; (iv) 1D random + $5$  ACS readout lines. The data were compressed to eight virtual coils. Quantitative results and representative frames are in Table \ref{Tab:1} and Figure \ref{Fig:1}, respectively. Not surprisingly, the performances of CF and SAKE* are similar because they essentially solve the same 2D problem, but CF converges in fewer iterations and has a significantly smaller memory footprint.

\begin{figure}[ht!]
	\includegraphics[width= 8.4cm]{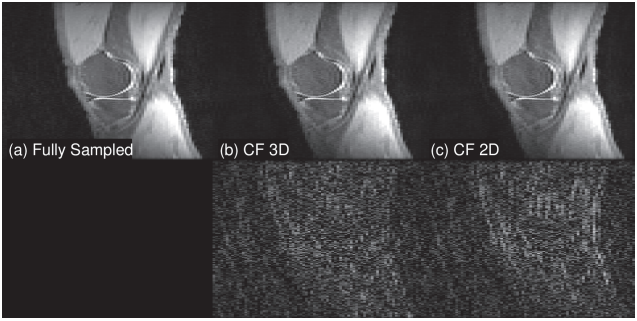}
	\centering	
	\caption{A representative slice from the 3D knee dataset.  SSoS image (top row); $10\times$ absolute error (bottom row). Columns are:  fully sampled image;  3D reconstruction via CF ($23.2$\,dB) and slice-by-slice reconstruction via CF ($22.8$\,dB). Compared to CF 3D, the error map of CF 2D has more structure.}
	\label{Fig:2}
\end{figure}

For 3D, we truncated a 3T knee dataset (downloaded from mridata.org) to $160k_x \times 80k_y \times 64k_z$, then retrospectively downsampled in $k_y$ and $k_z$ using 2D random sampling, with $15 k_y \times 15k_z$ ACS and fully sampled $k_x$. The data were compressed to four virtual coils for faster processing. A representative slice is shown in Figure \ref{Fig:2}. Since 3D CF reconstruction utilizes similarity and redundancy in three dimensions, with smaller degrees of freedom, it is able to outperform 2D CF reconstruction, which was separately applied to individual 2D slices. Due to prohibitive memory requirements, it was not feasible to extend SAKE's implementation to 3D.

\begin{figure}[h!]
	\includegraphics[width = 8.4cm]{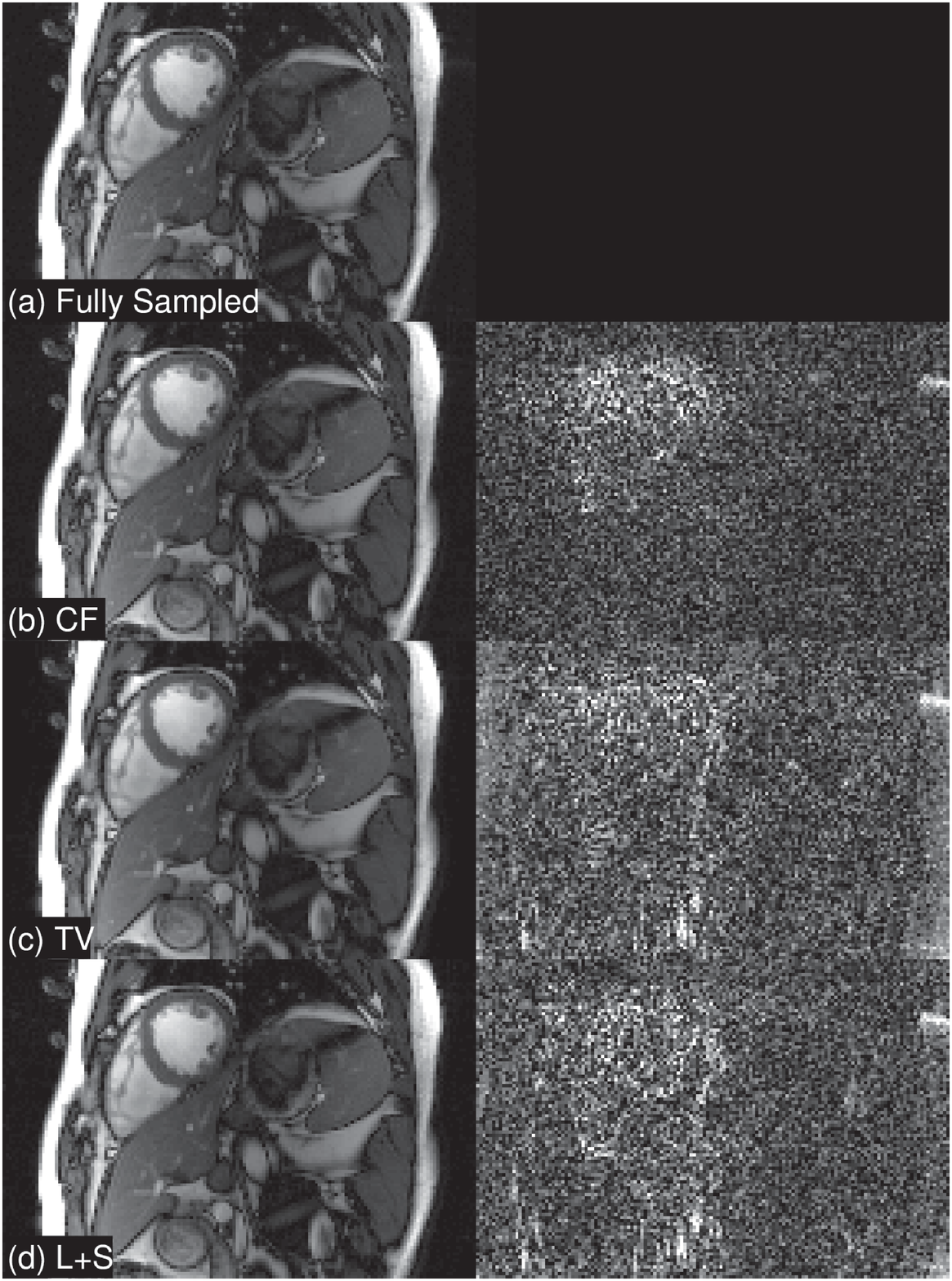}
	\centering
	\caption{A representative frame from 2D+t reconstruction. SSoS images (left) and $20\times$ absolute error (right).
	(a) fully sampled; (b) CF ($29.0$\,dB); (c) TV ($27.1$\,dB); (d) L+S ($26.8$\,dB). Compared to CF, the error maps of TV and L+S have more structure.}
	\label{Fig:3}
\end{figure}

For 2D+t, we retrospectively downsampled three 3T cardiac cine datasets at four different acceleration rates, R~$=4,6,8,$~and $10$ using a variable density sampling pattern~\cite{ahmad2015variable}. The data were compressed to four virtual coils for faster processing. We fine-tuned parameters for all methods with respect to the first dataset, then applied these parameters to other two datasets. We averaged the reconstruction SNR for the other two different datasets. Quantitative results and representative frames are in Table \ref{Tab:2} and Figure \ref{Fig:3}. We compare to SENSE-based techniques, as SAKE and LORAKS do not provide extension to these cases. The CF is consistently better than L+S \cite{otazo2015low} and TV \cite{lustig2007sparse} by  $1.1$ to $2.3$\,dB in terms of k-space SNR. With square-root sum of squared coils (SSoS), the margin is even larger.

\begin{table}[h!]
\centering
\begin{tabular}{|c|c|c|c| }
	\hline
	  & R=4 & R=6 & R=8\\
	\hline
	CF & \bf{22.4\,dB} & \bf{19.4\,dB} & \bf{15.8\,dB}\\
	 & 147.3\,s & 306.6\,s & 382.7\,s \\
	\hline
	SAKE & 20.5\,dB & 16.3\,dB & 13.7\,dB \\
	  & 301.2\,s & 320.7\,s & 293.7\,s\\
	\hline
	SAKE$^*$ & \bf{22.4\,dB} & \bf{19.4\,dB} & \bf{15.8\,dB}\\
	 & 593.5\,s & 772.2\,s & 722.8\,s\\
	\hline
	P-LORAKS & 22.3\,dB & 18.2\,dB & 14.4\,dB\\
	 & \bf{123.6\,s} &  \bf{158.6\,s} & \bf{156.2\,s}\\
	\hline 
	\end{tabular}
	\caption{2D k-space reconstruction SNR (dB) and time (s) comparison. Median time and average k-space SNR are computed across two datasets and four different sampling patterns. SAKE$^*$ denotes continuing SAKE past $200$ iterations until reaching CF SNR. }		
	\label{Tab:1}
\end{table}

\begin{table}[h!]
\centering
\begin{tabular}{|c|c|c|c|c| }
	\hline
	 & R=4 & R=6 & R=8 & R=10\\
	\hline
	CF & \bf{29.5} & \bf{28.2} & \bf{27.3} & \bf{26.3}\\
	\hline
	L+S & 27.1 & 26.4 & 25.5 & 25.1\\
	\hline
	TV & 27.2 & 26.1 & 25.1 & 24.1\\
	\hline
	\end{tabular}	
	\caption{2D+t k-space  reconstruction SNR (dB) averaged over two datasets. CF offers more than one dB advantage across all acceleration rates.}
	\label{Tab:2}
\end{table}

\section{Discussion and Conclusion}
Many parallel imaging approaches interpolate missing k-space points by solving a rank-deficient matrix completion. Choice of solution method can affect performance and memory requirements for this non-convex problem. CF provides an effective, memory-efficient solution and can subsume modeling choices found in existing GRAPPA-inspired methods. The memory requirement for CF processing (EIG + GD + ELS) is approximately the storage of the fully sampled k-space tensor $\mathbb{D}$. For example, for a 3D static double-precision complex k-space tensor with size $256 k_x \times 256 k_y \times 256 k_z \times 8 coils$ and a kernal size $10k_x \times 10k_y \times 10k_z \times 8coils$, a SAKE computation requires $2$ TB of memory, while CF only needs $2$ GB. 

Computation speed of CF can be further enhanced by using the following properties and processing steps.
\emph{Automatic filter size adaptation}: Many annihilating relationships can be efficiently and approximately captured by a small kernel. Thus, the CF computation can be sped up by adaptively progressing from small kernels to larger ones as iterations evolve. \emph{Automatic center to full k-space reconstruction}: Because the annihilation relationship holds for the entire k-space, we can apply CF for a small region of k-space and extract the null space, then enforce the estimated null space for the whole k-space. \emph{Highly parallizable}: All convolutions inside each iterative step are independent and thus highly parallizable. Thus, we can fully utilize multi-cluster, multi-core CPU, or GPU architectures to accelerate the processing.
\emph{Johnson–Lindenstrauss Lemma (JLL)}: The dimensionality of the null space and thus the size of the least squares problem can be further reduced by applying JLL in each iteration. In summary, simple conceptual framework, broad applicability, unifying nature, memory efficient computation, and a potential to further improve the computation speed make CF an attractive framework for MRI reconstruction.
		

%
%
\bibliographystyle{IEEEbib}
\bibliography{root.bib}

\end{document}